\documentclass[useAMS,usenatbib,usegraphicx,twocolumn]{mn2e}
\usepackage{times}
\usepackage{graphicx,graphics,epsfig}
\usepackage{amssymb,amsmath}

\def\araa{ARA\&A}
\def\mnras{MNRAS}
\def\apj{ApJ}
\def\aj{AJ}
\def\aap{A\&A}

\def\physrep{Phys. Rep.}

\newcommand{\vt}{{\vec{\theta}}}

\def\P{{\mathcal{P}}}

\def\HI{H~{\sc i}\ }

\newcommand{\bias}[1]{\mathsf{bias}[#1]}

\newcommand{\betahat}{\widehat{\beta}}
\newcommand{\betaobs}{\widehat{\beta}_{obs}}
\newcommand{\sehat}{\widehat{\mathsf{se}}}

\begin{document}

\title[The structure function of Galactic \HI opacity fluctuations]{The 
structure function of Galactic \HI opacity fluctuations on AU scales based on 
MERLIN, VLA and VLBA data}
\author[P. Dutta et al. (2013)]
{Prasun Dutta$^{1}$\thanks{Email: prasun@ncra.tifr.res.in}, 
Jayaram N. Chengalur$^{1}$, Nirupam Roy$^{2}$, W. M. Goss$^{3}$, 
\newauthor
Mihir Arjunwadkar$^{1}$, Anthony H. Minter$^{4}$, Crystal L. Brogan$^{5}$ and 
T. J. W. Lazio$^{6}$\\~\\
$^{1}$ National Centre for Radio Astrophysics, Post Bag 3, Ganeshkhind, Pune 411 007, India\\
$^{2}$ Max-Planck-Institut f\"{u}r Radioastronomie, Auf dem H\"{u}gel 69, D-53121 Bonn, Germany\\
$^{3}$ National Radio Astronomy Observatory, 1003 Lopezville Road, Socorro, NM 87801, USA\\
$^{4}$ National Radio Astronomy Observatory, P.O. Box 2, Green Bank, WV 24944, USA\\
$^{5}$ National Radio Astronomy Observatory, 520 Edgemont Road, Charlottesville, VA 22903, USA\\
$^{6}$ Jet Propulsion Laboratory, California Institute of Technology, Pasadena, CA 91109, USA
} 

\date{Accepted yyyy month dd. Received yyyy month dd; in original form yyyy month dd}
\pagerange{\pageref{firstpage}--\pageref{lastpage}} \pubyear{2013}
\maketitle 
\label{firstpage}

\begin{abstract}
We use MERLIN, VLA and VLBA observations of Galactic \HI absorption towards 
3C~138 to estimate the structure function of the \HI opacity fluctuations at 
AU scales. Using Monte Carlo simulations, we show that there is likely to be a 
significant bias in the estimated structure function at signal-to-noise ratios 
characteristic of our observations, if the structure function is constructed 
in the manner most commonly used in the literature. We develop a new estimator 
that is free from this bias and use it to estimate the true underlying 
structure function slope on length scales ranging $5$ to $40$~AU. From a power 
law fit to the structure function, we derive a slope of $0.81^{+0.14}_{-0.13}$, 
i.e. similar to the value observed at parsec scales. The estimated upper limit 
for the amplitude of the structure function is also consistent with the 
measurements carried out at parsec scales. Our measurements are hence 
consistent with the \HI opacity fluctuation in the Galaxy being characterized 
by a power law structure function over length scales that span six orders of 
magnitude. This result implies that the dissipation scale has to be smaller 
than a few AU if the fluctuations are produced by turbulence. This inferred smaller dissipation scale implies that the dissipation 
   occurs either in (i) regions with densities $\gtrsim 10^3 $cm$^-3$ (i.e. similar
   to that  inferred for "tiny scale" atomic clouds or (ii) regions with
   a mix of ionized and atomic gas (i.e. the observed structure in the
   atomic gas has a magneto-hydrodynamic origin).
\end{abstract}

\begin{keywords}
ISM: atoms -- ISM: general -- ISM: structure -- radio lines: ISM -- physical data and process: turbulence 
\end{keywords}

\section{Introduction}

The neutral atomic hydrogen(H~{\sc i}) component of the Galactic interstellar 
medium (ISM) is observed to have structure on a wide range of spatial scales. 
Studies over the last several decades have shown that this structure can be 
characterized by a scale free power spectrum on scales varying from a fraction 
of a parsec to hundreds of parsecs \citep{1983A&A...122..282C,1993MNRAS.262..327G,2001ApJ...561..264D,2000ApJ...543..227D,2009MNRAS.393L..26R}. This is 
generally understood to be the result of compressible fluid turbulence in the 
ISM. The turbulence is in turn believed to be generated by supernovae shock 
waves, spiral density waves in the disk etc. \citep[see, e.g.][for reviews]{2004ARA&A..42..211E,2004ARA&A..42..275S}. In addition to this parsec scale 
structure, fine scale structure on scales of tens of AU have also been 
observed in the atomic ISM \citep{1976ApJ...206L.113D,1996MNRAS.283.1105D,2001AJ....121.2706F,2005AJ....130..698B,2009AJ....137.4526L,2010ApJ...720..415S}. 
However, the connections, if any, between these AU scale structures and the 
structures observed at larger scales are not well understood. The existence of 
ubiquitous AU scale structure implies either that these structures are long 
lived and are in pressure equilibrium with the much lower (typically orders of 
magnitude smaller) density gas surrounding them, or that they are being 
continuously created \citep{2007A&A...465..431H}. Recent numerical simulations 
\citep{2006ApJ...643..245V,2006ApJ...652L..41N,  2007A&A...465..431H} suggest 
different mechanisms for generating and sustaining these AU scale structures. 
An alternative model proposed by \citet{2000MNRAS.317..199D} postulates that 
the observed fine scale structure is the projection of larger scale structures 
in the plane of the sky. In this picture, the AU scale structure would form 
part of the same scale free structure observed at parsec scales.

Here we use combined MERLIN\footnote{MERLIN: Multi-Element Radio-Linked 
Interferometer Network}, VLA\footnote{VLA: Very Large Array} and 
VLBA\footnote{VLBA: Very Long Baseline Array} observations to estimate the 
structure function of the \HI absorption towards 3C~138. Monte Carlo 
simulations show that the noise bias in the structure function, estimated from 
data with low signal to noise, is significant and is also scale dependent. We 
develop a new estimator that is free from this bias and use this to estimate 
the true underlying structure function on AU scales. Finally we compare the 
structure function that we measure on these small scales with that determined 
at larger scales.

\section{Statistical description of the optical depth fluctuations}
\label{sec:statis}

\subsection{Formalism}

The total optical depth in the \HI 21 cm spectral line towards a direction 
$\vt$ can be written as 
\begin{equation}
\tau(\vt)\ =\ \langle \tau(\vt) \rangle \ +\ \delta \tau(\vt) \,,
\label{eq:taust}
\end{equation}
where $\delta \tau(\vt)$ is the fluctuation about the mean optical depth 
$\langle \tau(\vt) \rangle$. The optical depth $\tau(\vt)$ is also a function 
of frequency. For notational simplicity, we do not explicitly show the 
frequency dependence.

Two related statistical measures widely used to quantify the properties of 
such fluctuations are the power spectrum and the structure function. Here we 
use the structure function defined as:
\begin{equation}
S_{\tau}(\vec{\phi})\ =\ \langle \left [ \tau(\vt+\vec{\phi}) -
  \tau(\vt)  \right ] ^{2} \rangle \,.
\label{eq:sfdef}
\end{equation}
The angular brackets in the above two equations denote the ensemble average 
over different realizations of the optical depth image. In practice, we have 
only one realization of the sky. The average is hence computed over different 
values of $\vt$ in the image assuming that the optical depth fluctuations are 
statistically homogeneous and isotropic. In such a case, the observed 
structure function depends only on the absolute value of the separation, viz. 
$\mid \vec{\phi} \mid$. The power spectrum of the optical depth fluctuations 
is defined as 
\begin{equation}
P_{\tau}(U)\ =\ \langle \mid \tilde{\delta \tau} (U) \mid ^{2}\rangle,
\label{eq:psdef}
\end{equation}
where $\tilde{\delta \tau} (U)$ is the Fourier transform of the optical depth 
fluctuations $\delta \tau({\vt})$ and $U$ is the inverse angular scale. In the 
case of scale invariant fluctuations, both the structure function and the 
power spectrum are power laws, viz. $S_{\tau}(\vec{\phi})\ =\ S_{0}\phi^{\beta}$ and $P_{\tau}(U)\ =\ A\, U^{\alpha}$, where $\alpha = -(\beta+2)$ \citep{1975ApJ...196..695L}.

\subsection{Measurement of the optical depth fluctuation}

We consider a situation where continuum radiation from an extended bright 
background radio source is being absorbed by a foreground \HI cloud. Observed 
intensity $I(\vt, \nu)$ towards a direction $\vt$ and at a frequency $\nu$ can 
be written as:
\begin{equation}
I(\vt, \nu) \ =\ I_{C}(\vt)\, e^{-\tau(\vt, \nu)},
\label{eq:line}
\end{equation}
where $I_{C}(\vt)$ is the radio continuum intensity distribution, and 
$\tau(\vt, \nu)$ is the optical depth in the \HI~21 cm line at the frequency 
$\nu$. If the observations are made over a sufficiently wide bandwidth, then 
the continuum intensity distribution $I_{C}(\vt)$ can be estimated from the 
observed intensities at frequencies at which $\tau(\vt, \nu) = 0$. 
Equation~(\ref{eq:line}) can then be inverted to obtain $\tau(\vt, \nu)$, from 
which in turn the structure function of the optical depth fluctuations can be 
estimated. This is the approach adopted by \citet{2000ApJ...543..227D} and 
\citet{2012ApJ...749..144R}. An alternative, Fourier space approach has been 
used by \citet[][and references therein]{2009MNRAS.398..887D} to estimate the 
power spectrum of the \HI intensity fluctuations. This approach is based on 
the fact that the primary observable in radio interferometry are the 
visibilities, which are Fourier transforms of the sky brightness distribution. 
In this approach, the optical depth fluctuation power spectrum is obtained by 
deconvolving the observed power spectrum of the fluctuations at the line 
frequencies with the power spectrum of the continuum fluctuations as estimated 
from the line free frequencies \citep[see][]{2010MNRAS.404L..45R}. For a 
continuum source with complicated geometry, this deconvolution can become 
impractical. This is the case here, where the background source (3C~138) has 
an elongated structure (of size $\sim 500$ milliarcsec) with complicated 
geometrical features. We hence use an image based approach to estimate the 
structure function of the optical depth fluctuations. A simple inversion of 
equation~(\ref{eq:line}) ignores the measurement noise, which is justified 
only when the observations have a large signal to noise ratio (henceforth 
SNR). In the next section, we examine at the effect of the measurement noise 
on the inferred structure function.

\subsection{ The effect of measurement noise on the estimated structure function}

\begin{figure*}
\begin{center}
\epsfig{file=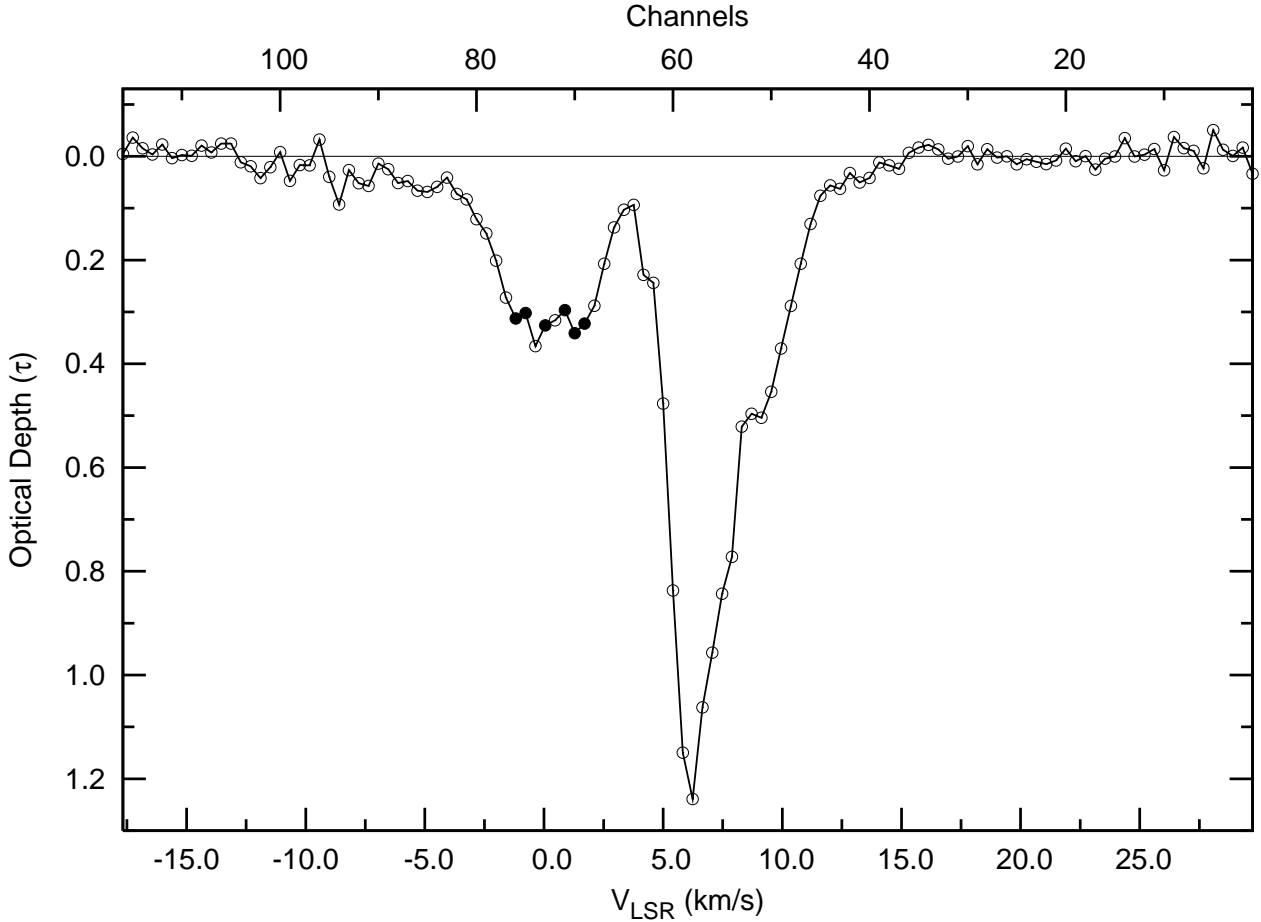,width=4.8in,angle=-90}
\end{center}
\caption{Average Galactic H~{\sc i} optical depth spectrum along the line of 
sight of 3C~138 from the combined MERLIN, VLA and VLBA data set is plotted 
against the LSR velocities. The corresponding channel numbers are marked on 
the top axis. The channels for which the structure function could be measured 
(see Sec.~\ref{sec:res}) are marked by filled circles in the velocity range 
$-1.2$ to $1.7$ km~s$^{-1}$.} 
\label{fig:spectra}
\end{figure*}

The sky brightness distribution inferred from the observed visibility data 
always have contribution from the measurement noise. This measurement noise 
affects both the continuum image and the image in each line channel (i.e, both 
$I(\theta, \nu)$ and $I_{C}(\theta)$), and this noise propagates into the 
estimated optical depth. Including this noise term, the measured optical depth 
can be written as:
\begin{equation}
\tau(\vt)^{(\rm obs)}\ =\ \langle \tau(\vt) \rangle \ +\ \delta
\tau(\vt)\ +\ \epsilon(\vt),
\label{eq:tauobs}
\end{equation}
where $\epsilon(\vt)$ is the measurement noise in the optical depth. The 
continuum  image is generally made by averaging over a large number of line 
free channels in the data cube. So, the noise in the continuum image is small 
compared to the noise in the line channels. If we assume that the noise in the 
continuum image is negligible, the mean and the standard deviation of 
$\epsilon(\vt)$ can be given as (see Appendix~\ref{app:bias})
\begin{eqnarray}
\langle \epsilon \rangle\ &=&\ \frac{1}{2}\, \left [  e^{ \tau (\vec{\theta})}\,
\frac{\sigma_{L}}{I_{C}(\vec{\theta})} \right ]^{2}  \nonumber  {\rm ~~ and}\\
\sigma_{\epsilon} \ &=&\ e^{ \tau (\vec{\theta})}\,
\frac{\sigma_{L}}{I_{C}(\vec{\theta})},
\label{eq:taunoise}
\end{eqnarray}
where $\sigma_{L}$ is the rms noise in the line image at the frequency of 
interest.  The average ``$\langle \rangle$'' mentioned here  is to be taken
over many realizations of the sky. The mean value of $\epsilon$ is
non-zero, i.e. the optical depth  measurement has a bias. Both the
mean and standard deviation of $\epsilon$  
depend on the true optical depth, the continuum emission and the 
measurement noise. All these three quantities are different for different pixels. Since observationally we have only one realization of the sky,
we can not measure and subtract the above-mentioned bias directly from the data. We discuss an alternate method to estimate the structure function below.

The structure function obtained from this optical depth 
image can be written as:
\begin{eqnarray}
S^{(obs)}_{\tau}(\vec{\phi})\ &=&\ \langle \left [ \tau^{(obs)}(\vec{\theta}) -
\tau^{(obs)}(\vec{\theta}+\vec{\phi}) \right]^{2}\rangle \nonumber \\
&=&\ S_{\tau}(\vec{\phi})\ +\ S_{\epsilon}(\vec{\phi})
+ {\rm cross\ terms},
\label{eq:sfobs}
\end{eqnarray}
where $S_{\epsilon}(\vec{\phi})$ is the structure function of the noise in the 
optical depth. Since the  bias in the optical depth measurement depends on the 
optical depth itself (equation~\ref{eq:taunoise}), the cross terms in 
equation~(\ref{eq:sfobs}) are non-zero. Thus, the structure function has a 
bias, which is different at different angular scales. In other words, the 
structure function measured in the presence of noise could have a shape that 
is different from that of the true structure function. The importance of this 
effect depends on the signal to noise ratio of the measurement. In the 
following sections, we briefly describe the observed data for 3C~138 and then 
the numerical simulations that we have carried out to quantify and correct for 
the effect of the bias.

\subsection{Description of the data}
Our analysis is based a combined data set obtained from MERLIN, VLA and VLBA 
observations\footnote{MERLIN observation date October 22 and 23, 1993; VLA 
project code AS0410 and TEST, observation date September 07 and 13, 1991; VLBA 
project code BD0026, observation date September 10, 1995.} of 3C~138. Details 
of the different data sets used to make this image, as well as the data 
analysis procedure followed, are presented by \citet{2012ApJ...749..144R}. To 
summarize, a spectral data cube was made from the combined calibrated uv-data 
using multi-scale CLEAN. The data cube has a spectral resolution of $0.4$ 
km~s$^{-1}$ per channel and a spatial resolution of $20$ milliarcsec (mas) 
with rms noise per channel about $20$ mJy. The source is unresolved in the VLA 
observations, so the VLA data allow us to accurately measure the total flux 
density. Further, the MERLIN, VLA and VLBA have overlapping baselines. The 
combined data set hence samples all angular scales between the total extent of 
the source ($\sim 500$ mas) and the largest angular scale probed by the VLBA 
($\sim 6$ mas). Fig.~\ref{fig:spectra} shows the average Galactic H~{\sc i} 
optical depth spectra towards 3C~138.

\section{Monte Carlo simulations}
\label{sec:mc}

To model the observations, we start by generating a simulated optical depth 
image. First, an image ($\delta   \tau^{(M)} (\vec{\theta})$) with zero mean 
Gaussian random fluctuations is generated with the input power law spectrum 
$P_{\tau}(U)$\footnote{ See \citet{2006ApJ...650..529W} for a detailed 
description of how to generate  Gaussian random fluctuations with a given 
power spectrum.}. While generating the image, we match the pixel size and 
total image size to the corresponding values in the actual data. We then 
modify the image by adding a fixed optical depth (i.e. the mean 
$\tau^{(M)}_{0}$)  to all pixels of the image. The resultant image is \\
$$\tau^{(M)} (\vec{\theta})\ =\ \tau^{(M)}_{0} +
\delta \tau^{(M)}  (\vec{\theta}) \,.$$

Next we use the continuum image from the observation and the model optical 
depth  image generated above ($\tau^{(M)} (\vec{\theta})$) to generate the 
line image, using equation~(\ref{eq:line}). To this line image, we then add 
measurement noise $N(\vec{\theta})$. This noise is generated from the same 
cumulative distribution function as that of the noise in the off-source pixels 
in the observed line image. Finally, we convolve the simulated line image with 
the same Gaussian restoring beam $B(\vec{\theta})$ as was used in the 
observations. This final model for the line image can be written as:
\begin{equation}
I^{(M)}(\vec{\theta})\ =\ \left[
  I_{C}(\vec{\theta})\, e^ { 
  -\tau^{(M)}(\vec{\theta}) }  +\, N(\vec{\theta}) \right]
\otimes B(\vec{\theta}) \,.  
\label{eq:modint}
\end{equation}
We then use $I^{(M)}(\vec{\theta})$ and $I_{C}(\vec{\theta})$ to invert 
equation~(\ref{eq:line}) and obtain the simulated optical depth distribution 
$\tau^{(S)}(\vt)$. The bias in the estimated structure function distribution 
can be derived by comparing the structure function measured from this 
simulated image with the known input structure function. A large number of 
realizations will allow an accurate determination of the bias. 

For running the simulations, we need three main input parameters: the 
amplitude and slope of the fluctuation power spectrum, and the mean value of 
the optical depth. Several previous studies \citep[e.g.][]{2010MNRAS.404L..45R,2000MNRAS.317..199D} have found that the power spectrum of \HI opacity 
fluctuations in the Milkyway, as well as in external galaxies, is well 
described by a power law with slope $\sim -2.6$ to $\sim -2.8$. Based on this, 
we assume that  $P_{\tau}(U) =\ A\, U^{\alpha}$, with $\alpha = -2.7$ and $A = 
1.0\times 10^{18}$ ($U$ in radians$^{-1}$). This value of the power spectrum 
amplitude $A$ is chosen so that the amplitude of the simulated structure 
function is comparable to the observed structure function (see below). 

Without any measurement noise, the observed mean optical depth is expected to 
be same as the mean optical depth used in the simulation. However, as 
described above (equation~\ref{eq:taunoise}), the optical depth measured in 
this way suffers from a bias, and hence differs from the true mean optical 
depth $\tau^{(M)}_{0}$. Specifically,
\begin{equation}
\langle \tau^{(S)}(\vt) \rangle\ =\ \tau^{(M)}_{0}  + \frac{1}{2}\,
  e^{ 2 \tau^{(M)}_{0}}\, \langle \left [
    \frac{\sigma_{L}}{I_{C}(\vec{\theta})} \right ]^{2} \rangle \,,
\label{eq:tmbias}
\end{equation} 
where $\langle \tau^{(S)}(\vt) \rangle$ is the optical depth averaged over the 
entire image. From our numerical simulations, we can determine $\langle 
\tau^{(S)}(\vt) \rangle$ and (partly as a test of the correctness of the 
simulations) compare these values to the known $\tau^{(M)}_{0}$.
 
\begin{figure}
\begin{center}
\epsfig{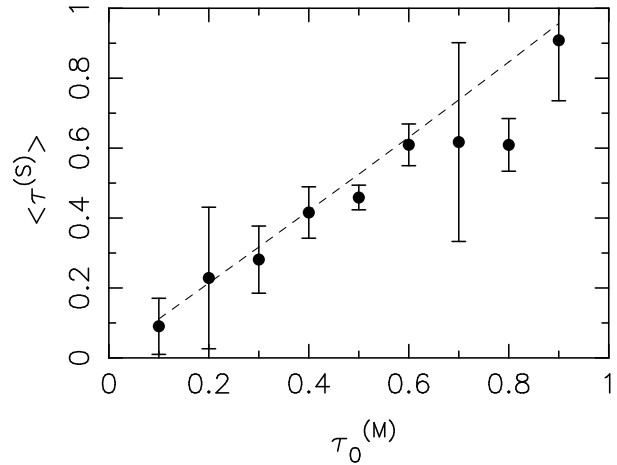}
\end{center}
\caption{The average optical depth $\langle \tau^{(S)}(\vt) \rangle$ as 
measured from the simulated optical depth images as a function of the true 
mean optical depth in the simulation $\tau^{(M)}_{0}$. The optical depth 
fluctuations are assumed to follow a power law with amplitude $A=1.0\times 
10^{18}$ and slope $\alpha=-2.7$. The averages and standard deviations are 
determined over 64 different realizations. The dashed line is the expected 
analytical relation (equation~\ref{eq:tmbias}).}
\label{fig:tau0av}
\end{figure}

Keeping $A$ and $\alpha$ fixed to the above values, $64$ different realizations 
of $\tau^{(S)}(\vt)$ images were generated for different $\tau^{(M)}_{0}$ 
values ranging between $0.1$ and $1.0$. The mean and the standard deviations 
of the averaged optical depth values from  these realizations are plotted 
against the  known true $\tau^{(M)}_{0}$ in Fig.~\ref{fig:tau0av}. The dashed 
line is the expected relation from equation~(\ref{eq:tmbias}) and can be seen 
to agree well with the results from the simulation. In the rest of the paper 
we have used this curve to determine the value of input $\tau^{(M)}_{0}$ to 
use with  a particular simulation.  

\begin{figure}
\begin{center}
\epsfig{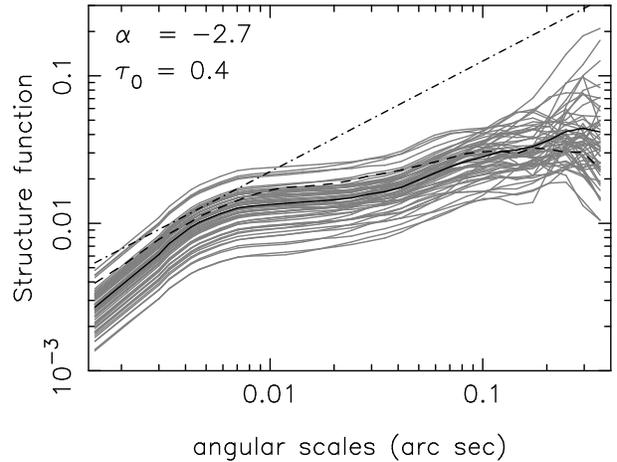}
\end{center}
\caption{Structure function estimated from 128 realizations with $\alpha=-2.7$ 
and $\tau_{0}=0.4$ is plotted against the angular scales (gray curves). Mean 
of these realizations is also plotted (black curve). The dot-dashed line has 
the same slope as the noise-free structure function. For comparison, the 
structure function estimated from one of the line channels with average 
optical depth $\sim 0.4$ from the data is also plotted.}
\label{fig:SFsim}
\end{figure}

To illustrate the effect of the bias in the measured structure function, we 
show in Fig.~\ref{fig:SFsim} the results from 128 simulation runs with the 
optical depth power spectrum amplitude fixed to $A = 1.0\times 10^{18}$  and 
slope $\alpha = -2.7$ as before and the mean optical depth set to 
$\tau^{(M)}_{0}= 0.4$. For each simulated image we estimate the structure 
function of the optical depth fluctuations. These structure functions are 
plotted in Fig.~\ref{fig:SFsim} as gray lines. The average structure function 
of the 128 realizations is also shown as a dark continuous line. For 
comparison, the structure function computed from a single line channel of the 
actual data cube is shown as a dashed line and the true underlying structure 
function (which has slope $\beta = -(2+\alpha)= 0.7$) is shown as a dot dashed 
line. The steep rise in the measured structure function at angular scales 
below  $0.01$ arc~sec reflects the fact that the pixel values are highly 
correlated for separations smaller than the resolution. The large scatter in 
the different structure functions at large angular scales is due to the fact 
that there are only few measurements at large separations. At intermediate 
scales, the measured structure function is consistent with a power law, but 
with a shallower slope than the true structure function. This effect 
illustrates the bias that was discussed in the previous section. As can be 
seen, the structure function estimated from the observed data is well matched 
by the structure functions estimated from the simulated data.

\subsection{An unbiased estimator for the slope of the structure function}

In this section we use the simulation procedure described above to construct 
an unbiased estimator $\hat{\beta}$ for the  structure function slope $\beta$. 
We start by noting from equation~(\ref{eq:taunoise}) that the bias in the 
observed optical depth is half of its  variance. Hence, the bias is reduced if 
the structure function is estimated from the pixels in the optical depth image 
with a high SNR. Restricting the measurement to only high SNR pixels, on the 
other hand, will also reduce the number of usable pixels. This procedure could 
also result in other more complicated biases by selectively sampling only part 
of the entire image. To determine the optimum value of SNR as a cutoff, we 
have apply different SNR cutoffs to the simulated optical depth images, and 
investigated which cutoff allows us to recover the input structure function. 

\begin{figure}
\begin{center}
\epsfig{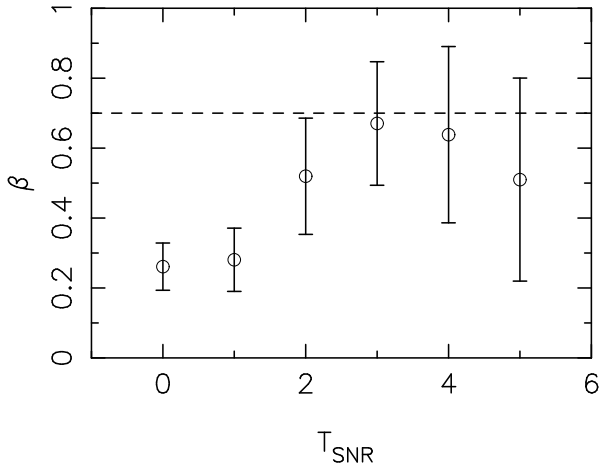}
\end{center}
\caption{Mean value and the standard deviation (as error bars) of the best fit 
Galactic H~{\sc i} structure function slope estimated from the simulated data 
is plotted for different signal to noise thresholds ($T_{SNR}$) in the optical 
depth values. The dashed horizontal line corresponds to an input slope of 
$0.7$. See the text for details.}  
\label{fig:snrfig}
\end{figure}

\begin{figure}
\begin{center}
\epsfig{file=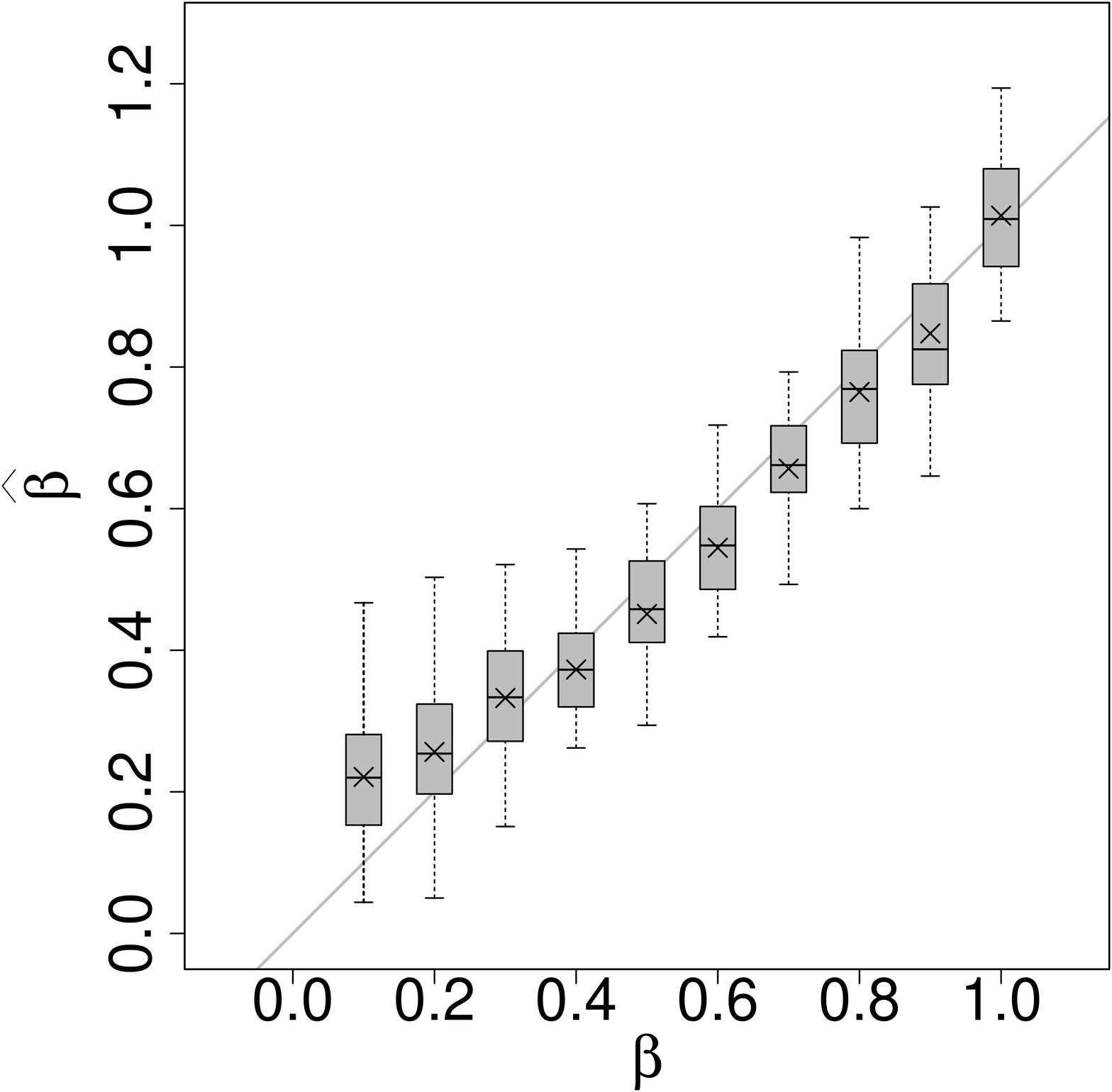,width=2.8in,angle=0}
\end{center}
\caption{Boxplots representing the distributions of $\betahat$ for different 
values of input $\beta$. Data boundaries (excluding the outliers) are marked 
by the bar ends, while the box boundaries mark 25\% and 75\% quantiles. The 
mean (crosses) and median (thick solid horizontal lines) from the points 
excluding the outliers are also shown. The number of data 
realizations for which $\betahat$ could be estimated are [59, 53, 48, 50, 42, 
46, 42, 35, 31, 27] (left to right) for the boxplots with input $\beta$ values 
from 0.1 to 1.0 respectively.}
\label{fig:betabias}
\end{figure}

As before, we set the optical depth fluctuation power spectrum amplitude and 
slope to $A = 1.0\times 10^{18}$ and $\alpha = -2.7$ respectively, and the 
mean optical depth to be $\tau^{(M)}_{0}= 0.4$. As discussed in more detail 
below, we chose $\tau^{(M)}_{0}= 0.4$ in order to maximize the number of 
pixels in the final image with SNR greater than the cutoff. We then compute 
the structure function from these images, using only pixels above a given 
threshold SNR value ($T_{SNR}$). Next, we fit a power law ( $S_{0}\, 
\phi^{\beta}$) in the angular scale range of $0.01$ to $0.05$ to the structure 
function estimated from each simulation and estimate the value of $\beta$ for 
each run. We exclude the  values of $\beta$ for which the fractional error 
from the fit is larger than $20\%$ (i.e. signal to noise less than 5). The 
mean and standard deviation are then calculated over the remaining 
realizations of the $\beta$ values. We repeat this procedure for different 
values of $T_{SNR}$. Fig.~\ref{fig:snrfig} shows the mean and standard 
deviation of the recovered $\beta$ values for different values of $T_{SNR}$. 
The dashed line corresponds to the true value of $\beta$. 
Naively, higher $T_{SNR}$ is expected to improve the accuracy in the estimate of $\beta$.  However, since high optical depth pixels also correspond to low SNR, the number of high optical depth pixels available to estimate $\beta$ decreases as the $T_{SNR}$ increases. This leads to a flattening of the structure function (or equivalently a decrease in $\beta$) as $T_{SNR}$ is increased.  As can be seen, at 
$T_{SNR} = 3$ the recovered $\beta$ is close to the true one; for higher and 
lower values of $T_{SNR}$ the recovered $\beta$ is offset from the true value. 
So at least for an input structure function slope of $0.7$, a cutoff value of 
$T_{SNR} \sim 3$ is optimal. To check whether it is optimal for other input 
structure function slopes, the same procedure is repeated for different values 
of $\alpha$ , viz.  $-3.0 < \alpha <-2.1$ (i.e., $1.0 > \beta > 0.1$). The 
results are summarized in Fig.~\ref{fig:betabias}, where the distribution of 
$\betahat$ for each input $\beta$ is shown in the form of a boxplot. The solid 
horizontal line (cross) inside each boxplot represents the median (mean) of 
the distribution of $\betahat$. For $\beta \gtrsim 0.4$, using a cutoff of 
$T_{SNR} = 3$ results in a value of $\betahat$ that is close to $\beta$. For 
lower values $\beta$, even with a cutoff $T_{SNR} = 3$, $\betahat$ is 
systematically offset from $\beta$.  In the range $0.4 \lesssim \beta \lesssim 0.9$, the mean value of  
$\betahat$ is offset from $\beta$ by a constant, viz. $-0.02$. So the final 
step in determining an unbiased value of the structure function slope would be 
to subtract this constant from $\hat{\beta}$. Apart from estimating $\beta$, 
we would also like to know the confidence intervals around the estimated 
$\beta$. We return to this issue after we have applied this estimator to the 
actual data.

The procedure discussed above leads to an unbiased estimate of the structure 
function slope. Even after using only the high signal to noise ratio pixels, 
the amplitude of the structure function, however, still has a bias, which 
cannot be easily estimated. This noise bias is positive. So, although we can 
not estimate the amplitude of the $S_{\tau}(\theta)$, we can set an upper 
limit to the structure function amplitude.

\section{Application to the 3C~138 data}
\label{sec:res}

\begin{figure}
\begin{center}
\epsfig{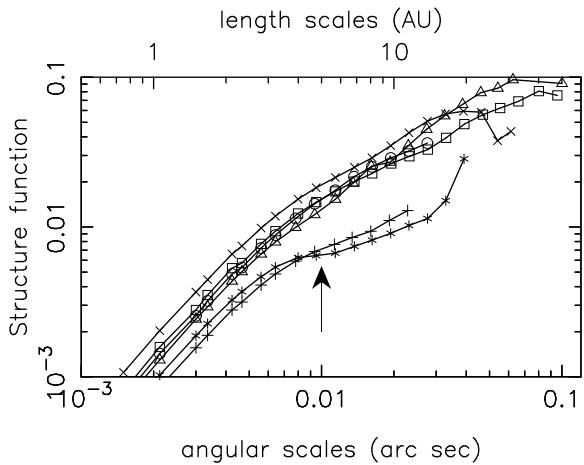}
\end{center}
\caption{Structure function of Galactic H~{\sc i} estimated using the pixels 
with three sigma sensitivity in optical depth for the six channels (69, 70, 
71, 73, 75, 76). The top axis also shows the corresponding length scales 
assuming that the distance to the \HI is 500~pc. The instrumental resolution 
results in a steepening in the structure function at angular scales smaller 
than $0.01$ arc seconds. This angular scale is marked by an arrow in the 
figure.}
\label{fig:SFres}
\end{figure}

Guided by the results discussed in Sec.~\ref{sec:mc}, we used an SNR cutoff 
$T_{SNR} = 3$ to estimate the structure function for all channels in the 
observed data cube with average optical depth lying between $0.2$ and $0.4$. 
This restriction in the mean optical depth occurs for the following reason. 
Since the noise in the optical depth image increases exponentially with 
optical depth (see equation~\ref{eq:taunoise}), the signal to noise ratio is 
low for pixels with either large or small optical depths. Consequently, 
channels with large or small average optical depth have very few pixels that 
lie above the $T_{SNR} = 3$ cutoff and generally sample a very restricted 
region of the image. The net result is that too little data are left to make a 
good estimate of the structure function. 

For spectral channels with mean optical depth between $0.2$ and $0.4$, the 
structure function was computed. For a total of 6 channels the structure 
function could be fit using a power law of the form $S_{\tau}(\phi)\ =\ S_{0} 
\phi ^{\beta}$ with an uncertainty in the slope of less than 20\% 
(Fig.~\ref{fig:SFres}). All of these channels are at velocities corresponding 
to local H~{\sc i}. The data cube also shows absorption at velocities 
corresponding to the more distant \HI with average optical depth values 
between $0.2$ and $0.4$. However in these channels the number of pixels that 
remained after using an SNR cutoff of three was too small to estimate the 
structure function. For the six channels for which the structure function 
could be estimated, the structure function slope was estimated using the 
unbiased estimator $\betahat$ introduced in Sec.~\ref{sec:mc}. As discussed in 
Sec.~\ref{sec:mc}, the measured amplitude of the structure function $S_{0}$ is 
an upper limit to the amplitude of the true structure function amplitude.

\begin{figure}
\begin{center}
\epsfig{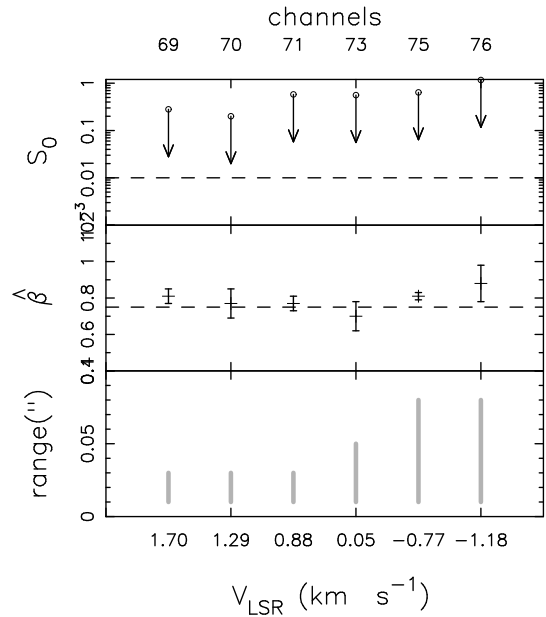}
\end{center}
\caption{Top panel shows the upper limit of the structure function amplitude 
estimated from six channels (channel numbers and the corresponding LSR 
velocities are given at the top and  bottom of the figure respectively) in the 
data cube. Points with error bars in the middle panel indicate the structure 
function slope and the gray bars in the bottom panel show the range of 
angular scales over which the fit could be done. Horizontal lines in the upper 
two panels indicates the best fit value from \citet{2000MNRAS.317..199D}.}
\label{fig:figres}
\end{figure}

Fig.~\ref{fig:figres} shows the resultant values for $S_{0}, \hat{\beta}$ and 
the range of angular scales over which the power law fit was done for the $6$ 
frequency channels. The top panel of this figure shows the resultant upper 
limits to the true structure function amplitudes. The middle panel in the same 
figure shows the slopes of the structure function and the gray bar in the 
bottom panel gives the range of angular scales for which the estimate could be 
made. The LSR velocities corresponding to these channels are given in the 
bottom axis of the plot. The dashed lines in the upper two panels show the 
amplitude and slope of the optical depth structure function as estimated on 
parsec scales \citep{2000ApJ...543..227D}.

All the six channels used to estimate the structure function are shown as 
filled circles in Fig. 1. For the channels (69, 70, 71), a power law can be 
fit over a small range of angular scales (0.01 to 0.03), whereas for the 
channels (75, 76), the range is comparatively broader (0.01 to 0.08). 
Following \citet{2001AJ....121.2706F} we adopt a distance of $500$ pc for the 
\HI that gives rise to the absorption. At this distance, the angular range of 
the fit corresponds to a linear range of $5$ to $40$ AU. The angular range 
used for the fit does not exceed a factor of 10 for any of the channels. As 
such, while a power law  is consistent with the structure function determine 
from the data, from the data we cannot prove that the structure function has a 
power law form. Other functional forms may also provide a good fit. 

\begin{figure}
\begin{center}
\epsfig{file=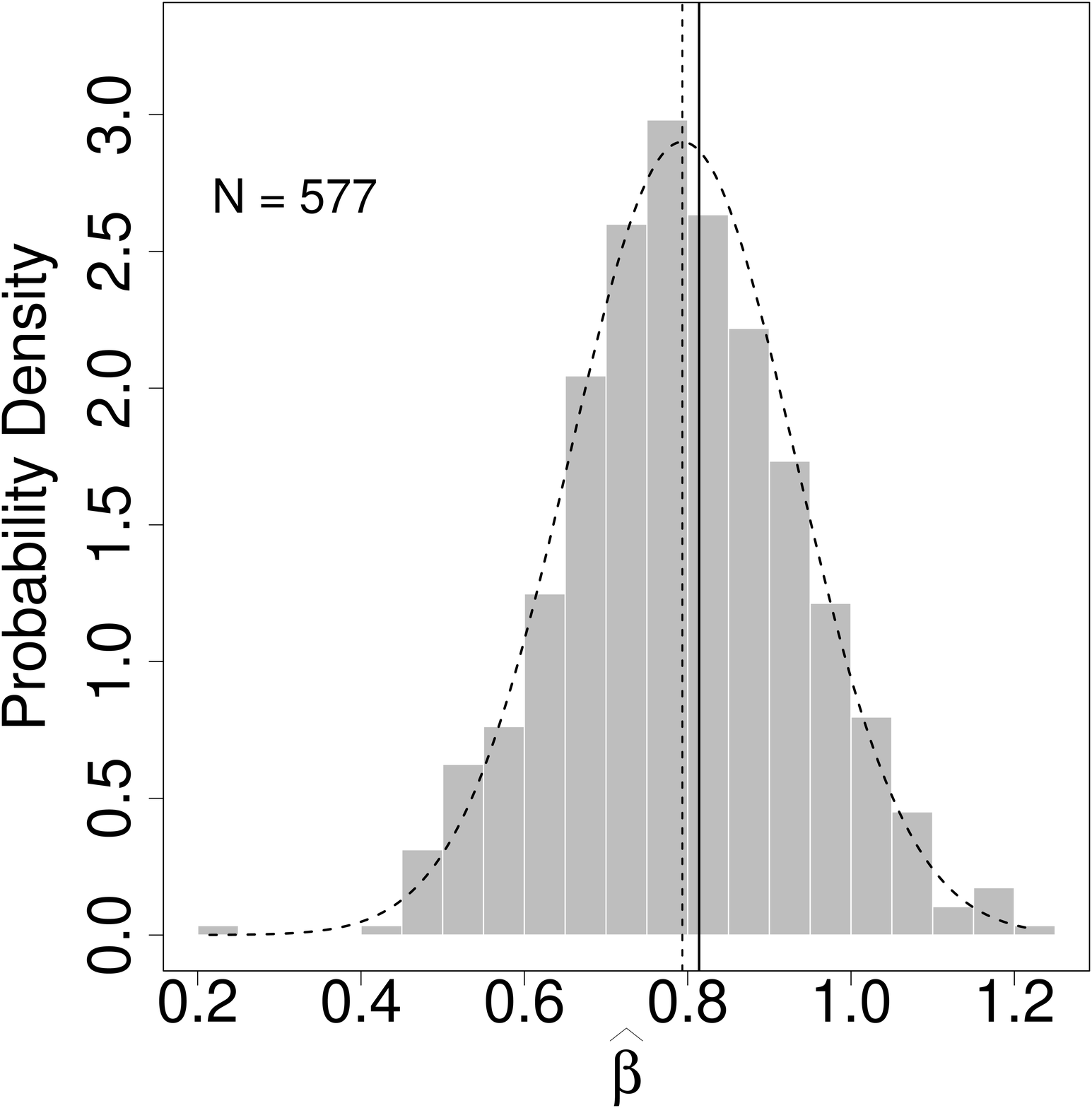,width=2.8in,angle=0}
\end{center}
\caption{Histogram showing the probability density of $\betahat$ estimated 
with an input $\beta = 0.81$ (input $\beta$ is indicated with a solid vertical 
line). The dashed vertical line indicates the mean of the histogram. The mean 
(dashed vertical line) and median are similar. For comparison, a Gaussian 
function with same mean and standard deviation of the data is also 
over-plotted (dashed bell-shaped curve). The total number of realizations used 
to produce the histogram is $N=577$.}    
\label{fig:histbeta} 
\end{figure}

In case of the six channels under discussion (see Fig. 1 for the channels 
analysed), the recovered slope is similar. The mean value of these six 
observational estimates of $\beta$  is $\betaobs = 0.79$ with a standard 
deviation of $0.06$. The error bars shown in Fig.~\ref{fig:figres} correspond 
to the error estimates for the power law fit. The estimated $\beta$ is well 
within the range for which the Monte Carlo simulations show that an unbiased 
recovery of the slope is possible. As discussed in Sec.~\ref{sec:mc}, in this 
range the estimator $\hat{\beta}$ has a small remaining bias of $-0.02$. 
Accounting for this, the final mean estimated slope is $0.81$. 

Given the estimated value of the structure function slope, we now turn to the 
question of the appropriate confidence interval for this estimate. 
Determination of the confidence intervals on the estimated $\betahat$ requires 
knowledge of the probability distribution of the estimator. To get some 
information on the probability distribution, we simulated a large number 
($2048$) of images with input $\beta = 0.81$ and estimate $\beta$ using the 
same procedure as used for the actual data. We denote the estimated standard 
deviation of this sample by $\sehat$. Given $\sehat$ there are two 
prescriptions for $(1-\gamma)$ confidence intervals ($0 \le \gamma \le 1$) 
based on the following two different assumptions: 
\begin{enumerate}
\item
Assuming that the distribution of $\widehat{\beta}$ is Gaussian with mean 
$\overline{\beta}$ and estimated standard deviation $\sehat$, the $(1-\gamma)$ 
confidence interval on $\overline{\beta}$ is $$\overline{\beta} \pm 
z_{\gamma/2} \times \sehat,$$ where $z_{x} = \Phi^{-1}( 1 - x )$ for $0 \le x 
\le 1$, and $\Phi(x) = 0.5\, (1 + \mbox{erf}(x/\sqrt{2}))$ is the cumulative 
distribution function for the standard (i.e., zero mean and unit standard 
deviation) Gaussian distribution. We call this the \emph{Gaussian-based} 
confidence interval.
\item
Assuming that the distribution of $\widehat{\beta}$ is unimodal but not 
necessarily symmetric (or Gaussian), the $(1-\gamma)$ confidence interval on 
$\beta$ is defined by the $\gamma/2$ and $(1-\gamma/2)$ numerical quantiles of 
this sample. We call this the \emph{quantile-based} confidence interval. In 
the case when $\betahat$ has a Gaussian distribution, these two prescriptions 
are equivalent. 
\end{enumerate}

Fig.~\ref{fig:histbeta} shows the histogram for the estimates of $\betahat$. 
The mean value (excluding five potential outliers not shown in the figure) of 
this distribution (the dashed vertical line) is $0.793$, close to the input 
$\beta$. The standard deviation is $\sehat=0.138$. The dashed curve in 
Fig.~\ref{fig:betabias} shows a Gaussian probability density function with the 
same mean and standard deviation as the data (excluding the outliers). The 
confidence intervals using both methods are given in Tab.~\ref{tab:int}. As 
can be seen, both methods provide similar confidence interval of 
$0.81^{+0.14}_{-0.13}$ ($1 \sigma$) on the structure function slope for 
spatial scales of $5$ to $40$~AU.

\begin{table}
\begin{tabular}{c|c|c}
  \hline 
  $\gamma$ & {\rm Quantile-based} & {\rm Gaussian-based} \\
  & & \\
  \hline \hline 
  $0.3173$ ($1\sigma$) & $0.68 - 0.95$ & $0.68 - 0.95$ \\
  $0.0455$ ($2\sigma$) & $0.54 - 1.08$ & $0.54 - 1.09$ \\
  $0.0027$ ($3\sigma$) & $0.48 - 1.18$ & $0.40 - 1.23$ \\
  \hline
\end{tabular}
\label{tab:int}
\caption{Quantile- and Gaussian-based confidence intervals for $\beta$ at the 
$1\sigma$, $2\sigma$, and $3 \sigma$ levels. Column (1), (2) and (3) gives the 
value of $\gamma$ (see text) and the corresponding Quantile- and 
Gaussian-based confidence intervals.}
\end{table}

\section{Discussion and conclusion}

\citet{2000ApJ...543..227D} and \citet{2009MNRAS.393L..26R} have estimated the 
H~{\sc i} opacity fluctuation power spectra and/or the structure function in 
the ISM of our Galaxy. \citet{2000ApJ...543..227D} used a very similar image 
based technique to estimate the opacity fluctuation structure function towards 
Cas~A. Those authors have not made any correction for the bias, which is 
justifiable in their case because the brightness of Cas~A is expected to make 
the bias small. This small bias in case of Cas~A is confirmed by the fact that 
a completely different visibility based approach (which does not suffer from 
this bias) gives a structure function slope that is in excellent agreement 
with the image plane base approach \citep{2009MNRAS.393L..26R}. Interestingly 
the current measurement of the structure function slope $\beta = 
0.81^{+0.14}_{-0.13}$ ($1 \sigma$, Gaussian-based) at length scale $5$ to $40$ 
AU is consistent with that observed towards Cas~A ($\beta = 0.75\pm0.25$) over 
almost six orders of larger scales (i.e. $~\sim 0.02$ to $\sim 4$ pc). The 
amplitude from the power law given by \citet{2000MNRAS.317..199D} (indicated 
in the top panel of Fig.~\ref{fig:figres} by a dashed line) is also consistent 
with the upper limit derived here.

\citet{2012ApJ...749..144R} also used the same data that are used here to 
estimate the structure function using a similar technique. The important 
difference is that \citet{2012ApJ...749..144R} used two methods: (1) no cutoff 
in optical depth and (2) a two sigma cutoff based on the optical depth signal 
to noise. This later choice is in contrast to the three sigma signal to noise 
used in the present analysis. Based on this lower signal to noise cutoff, 
\citet{2012ApJ...749..144R} were able to estimate the structure function over 
a larger range of angular scale (more than a decade) with 36 spectral 
channels. On the other hand, as mentioned in Sec. 4, the three sigma cutoff 
restricts the analysis to only six spectral channels. The structure function 
can then be estimated only over a smaller range of angular scales. However, 
the advantage of the current analysis, demonstrated using the numerical 
simulations described in Sec. 3, is that the structure function estimator with 
the three sigma cutoff is free from the scale dependent bias arising from the 
correlated noise. Using a three sigma cutoff is crucial to obtain an unbiased 
estimator; hence the present estimate of the structure function slope is more 
reliable. The current best fit value of $0.81^{+0.14}_{-0.13}$ is consistent 
with the \citet{2012ApJ...749..144R} value of $0.33\pm0.07$ at roughly a three 
sigma level. In future, these improved techniques can be applied to data with 
higher signal to noise in order to estimate the power spectra over a larger 
range of angular scales and a larger range of velocity channels.

\subsection{Turbulence dissipation scale}

As mentioned above, our fit of a power law model for the structure function is 
over a very small range of angular scales (order of ten). This functional form 
(viz. a power law) is motivated by the theoretical understanding that fine 
scale structure is expected to be the result of ISM turbulence and the fact 
that the structure function on much larger scales is observed to have a power 
law form. However, because of the small range of angular scales that we are 
restricted to in the current analysis, we can not {\it establish} that the 
structures at these scales actually follow a power law. Nonetheless, we do 
establish that the data are consistent with AU scale structures having a 
structure function that has the same slope as that observed on much large 
scales. In case of turbulence generated structures in a medium, the structure function assumes a power law at scales lying between the scale of  energy input (i.e, the driving scale) and  the scale at which energy is taken out (i.e, the dissipation scale). If the structures we see here are generated by turbulence, the turbulence 
dissipation scale should be smaller than $\sim 5$ AU. The dissipation scale 
for \HI in the ISM is discussed in detail by \citet{1998MNRAS.294..718S}. For 
a largely neutral gas with a thermal velocity dispersion of $\sim 1$ 
km~s$^{-1}$, and a Kolmogorov like turbulence, the Reynolds number $R_{e}$ is 
given by $R_{e} = 3\times 10^{4}\, \frac{L}{10\, {\rm pc}}\, \frac{v}{10\, 
{\rm km\, s}^{-1}}\, \frac{n}{1\, {\rm cm^{-3}}}$. Here, $v$ is the turbulent 
velocity dispersion, $n$ is the average density at the energy dissipation 
scale, and $L$ is the energy injection scale for turbulence. The dissipation 
scale $l_{d}$ in such a case is given by $l_{d} = L R_{e}^{-3/4}$. Typically 
$v$ is $\sim 5$ km~s$^{-1}$ for the neutral gas, and $L$ is $\sim 10$ pc for 
energy injection via supernova explosions.

Unfortunately, there is no strong or direct constraint on the density at the 
scales of our interest. If we assume the density to be $\sim 1 - 100\, {\rm 
cm^{-3}}$, similar to that of the diffuse cold H~{\sc i}, the dissipation 
scale is $\sim 50 - 1500$ AU, significantly larger than the $5$ AU scale probed in 
the current analysis. In this case, a higher Reynolds number (and hence a 
smaller dissipation scale) is possible if ionized gas is mixed with the 
H~{\sc i} \citep[e.g.][]{2005PhR...417....1B} and the turbulence is 
magnetohydrodynamic in nature. On the other hand, if the observed opacity 
fluctuations are directly related to the so-called ``tiny-scale atomic 
structure'' (TSAS), then the density can be $> 10^3\, {\rm cm^{-3}}$ 
\citep[e.g.][]{1997ApJ...481..193H}, and the dissipation scale for the neutral 
gas will be $\lesssim 5$ AU. The fact that we observe the structure function 
to be consistent with a power law at AU scales hence suggests that either the 
\HI at these scales is associated with ionized gas and thus the 
magnetohydrodynamic (MHD) turbulence is playing an important role, or that the 
observed fluctuations are related to TSAS with a few orders of magnitude 
higher density. 

Complimentary to the radio observations, optical/UV/IR studies have
also been used to trace small scale structures, though these different tracers may not necessarily trace the same ISM phase. Interstellar
absorption towards binary or common proper motion systems are used to
probe $\lesssim 2500$ AU scale. Broadly these observations suggest that
while warm gas (traced by, e.g. Ca~{\sc ii} absorption) does not show much
structures at these scales, small scale structures are relatively
common in colder gas (traced by, e.g. Na~{\sc i} absorption) at similar
scales (e.g. \citet{2003ApJ...591L.123L}, \citet{1994ApJ...436..152W}, \citet{1996ApJ...473L.127W}). \citet{2000ApJ...543L..43L} and \citet{2003ApJ...591L.123L}  reported N~{\sc i} structure as small as $\sim 10-20$ AU based on observed
temporal variation due to proper motion, where as \citet{2003A&A...401..215R} reported column density structures at similar scales (i.e. $\sim 10$
AU) from CH and CH$^+$ observations. Similarly, \citet{2004A&A...422..483A} have
also reported sub-pc scale molecular structure in multiple tracers.
\citet{2007ASPC..365...59G} used optical and radio observations of the Pleiades
reflection nebula to reveal a power law power spectrum of optical dust
reflection and H~{\sc i} radio emission with a power law index of -2.8 over
more than 5 order of magnitude range of scale down to few tens of AU.
On the other hand, \citet{2007ASPC..365...59G} have reported Spitzer
observations of the Gum nebula, for which the power spectrum of the
surface brightness has a power law with an index of $\alpha = -3.5$ at
millipersec scales, but $\alpha = -2.6$ at $> 0.3$ pc scales. Overall however, the
optical/UV/IR observations are consistent with the presence of
significant self-similar small scale structures in the ISM.

To summarize, we have used Monte-Carlo simulations to show that a direct 
estimation of the structure function slope from the optical depth image 
towards 3C~138 is contaminated by a scale  dependent bias. We further show 
that this bias can be overcome by  using only those pixels  with SNR greater 
than three to estimate the structure function. Using this prescription, we 
have estimated the structure function for absorption from the local gas 
towards 3C~138, and find that the structure function has a slope of 
$0.81^{+0.14}_{-0.13}$ for length scales from $5$ to $40$~AU. This value of the 
slope agrees within the error bars with the structure function slope measured 
on scales that are almost six orders of magnitude larger. The same power law 
slope extending down to scales of a few AU implies that, if these structures 
are produced by turbulence, the dissipation scale of the turbulence needs to 
be smaller than a few AU. Finally, we argue that such smaller dissipation 
scale is consistent with the theoretical models if either the density of these 
tiny scale structures is more than a few times $> 10^3\, {\rm cm^{-3}}$, or 
the H~{\sc i} turbulence is governed by MHD processes via fractional 
ionization and the presence of magnetic field.

\section*{Acknowledgments}

The authors are grateful to K. Subramanian, A. Deshpande, N. Kanekar, S. 
Bharadwaj and S. Bhatnagar for useful discussions. This paper reports results 
from observations with MERLIN, VLA and VLBA. MERLIN is a National Facility 
operated by the University of Manchester at Jodrell Bank Observatory on behalf 
of PPARC/STFC. The National Radio Astronomy Observatory is a facility of the 
National Science Foundation operated under cooperative agreement by Associated 
Universities, Inc. PD would like to acknowledge the DST - INSPIRE fellowship 
[IFA-13 PH-54 dated 01 Aug 2013] used while doing this research. NR 
acknowledges support from the Alexander von Humboldt Foundation and the Jansky 
Fellowship of the National Radio Astronomy Observatory. Part of this research 
was carried out at the Jet Propulsion Laboratory, California Institute of 
Technology, under a contract with the National Aeronautics and Space 
Administration.

\bibliographystyle{mn2e}

\appendix

\section{The Bias and Variance of the estimated optical depth}
\label{app:bias}

\begin{figure}
\begin{center}
\epsfig{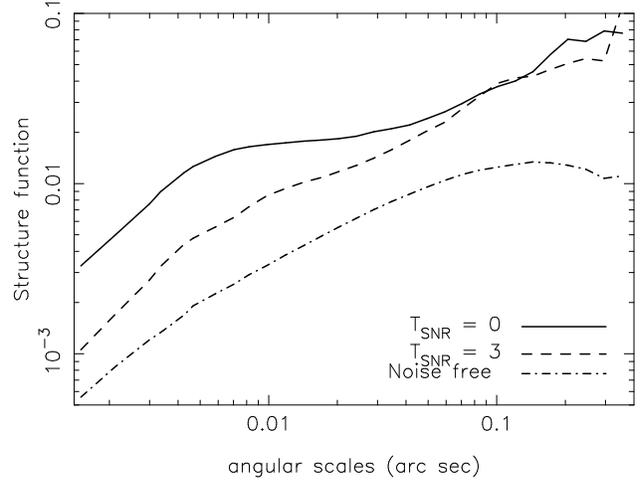}
\end{center}
\caption{Structure function estimated from one of the 128 realizations of the 
simulations described in section 3 is plotted for no signal to noise ratio 
cutoff (T$_{\rm SNR} = 0$, solid line) and a cutoff of $3\sigma$ (T$_{\rm SNR} 
= 3$, dashed line). The underlying noise-free structure function is shown 
using the dot dashed line. The structure function measured after using a 
cutoff of $3\sigma$ has the same slope as the true structure function. The 
measured structure function is however biased to higher amplitudes than the 
true structure function.}
\label{app:fig1}
\end{figure}

The observed optical depth ($\tau^{(obs)}$) at a frequency $\nu$ is given by
\begin{equation}
\tau^{(obs)}(\nu) = -\  \log \left [ \frac{I(\nu)}{I_{C}} \right ] \,,
\label{eqn:tauest}
\end{equation}
where $I(\nu)$ is the observed intensity at the frequency $\nu$ and $I_{C}$ is 
the observed continuum intensity. These observed intensities have associated 
measurement noise. Since the continuum image is obtained from averaging the 
emission from a large number of line free channels of the spectral cube, we 
assume that the noise in the continuum intensity is negligible compared to the 
noise in a single line channel. We hence have
\begin{equation*}
I(\nu) \ =\ I_{C}\, e^{-\tau(\nu)} \ +\ \epsilon,
\end{equation*}
where $\tau(\nu)$ is the true optical depth, and $\epsilon$ is the noise in 
the line image. Using this expression in Eqn.~(\ref{eqn:tauest}) gives
\begin{eqnarray}
\tau^{(obs)}(\nu) &=& -\  \log \left [ \frac{I_{C}\,
                  e^{-\tau(\nu)} \ +\ \epsilon}{I_{C}} \right ]\, \nonumber\\
             &=&\  \tau(\nu)\, - \log \left [ 1 + \frac{ \epsilon\,
                                   e^{\tau(\nu)}}{I_{C}} \right ] \,.
\label{eqn:tauestt}
\end{eqnarray}
The expected value of $\tau^{(obs)}(\nu)$ is
\begin{equation}
\langle \tau^{(obs)}(\nu)\rangle = \tau(\nu)\, - \Biggr \langle \log \left [ 1 + \frac{ \epsilon\,
                 e^{\tau(\nu)}}{I_{C}} \right ] \Biggr \rangle \,.
\label{eqn:tauave}
\end{equation}
Assuming that 
\begin{enumerate}
\item the signal to noise ratio is large, i.e. 
   $$\frac{ \epsilon\, e^{\tau(\nu)}}{I_{C}} \ll 1 \,,$$
\item the noise $\epsilon$ has a  Gaussian distribution with zero mean,
\end{enumerate}
we can use a Taylor expansion of the logarithm in Eqn.~(\ref{eqn:tauave}) to 
obtain
\begin{equation}
\langle \tau^{(obs)}(\nu)\rangle \ =\  \tau(\nu)\, + \frac{1}{2}\,\left
     [e^{\tau(\nu)}\, 
\frac{\sigma_{L}}{I_{C}} \right ]^{2},
\end{equation}
where $\sigma_{L}^{2} = \langle \epsilon^{2} \rangle$. The bias in the optical 
depth estimates is hence given by
\begin{equation}
\bias{\tau(\nu)}\ =\ \langle \tau^{(obs)}(\nu) \rangle -  \tau(\nu)\ =\  \frac{1}{2}\,\left
     [e^{\tau(\nu)}\, 
\frac{\sigma_{L}}{I_{C}} \right ]^{2} \,.
\end{equation}
The variance of $\tau^{(obs)}(\nu)$ is given by
\begin{equation}
\sigma^{2}_{\tau}\ = \ \langle (\tau^{(obs)}(\nu))^{2} \rangle -
\langle \tau^{(obs)}(\nu) \rangle^{2} \,.
\end{equation}
From Eqn.~(\ref{eqn:tauestt}) and the same assumptions as above, we can easily 
derive that
\begin{equation}
\langle (\tau^{(obs)}(\nu))^{2} \rangle = 
\tau^2(\nu) + \left [ e^{\tau(\nu)} \frac{\sigma_{L}}{I_{C}} \right]^{2}
 \left [ 1 + \tau(\nu) \right ].
\end{equation}
and hence that the variance $\sigma^{2}_{\tau}$ is
\begin{equation}
\sigma_{\tau}^{2}\ =\ \left
     [e^{\tau(\nu)}\, 
\frac{\sigma_{L}}{I_{C}} \right ]^{2}.
\end{equation}
Interestingly, the bias and variance are related, the bias being half of the 
variance. This implies that the bias in the optical depth image (and hence in 
the derived structure function) can be reduced by restricting the measurement 
to pixels with good signal to noise ratio. The estimator $\hat{\beta}$ for the 
slope of the structure function that we introduce in Sec.~\ref{sec:mc} is 
motivated by this observation.

The measurement noise will also affect the estimated amplitude of the 
structure function. Intuitively, the amplitude of the structure function 
measured from a noisy image is expected to be higher than the amplitude of the 
true structure function. We have used detailed analytical calculations to 
verify that, under the assumptions listed above, the measured structure 
function at the angular scales of interest is a scaled version (with scale 
factor greater than $1$) of the noise free structure function. The Monte Carlo 
simulations presented in the paper (see Fig.~\ref{app:fig1}) also points to 
such a multiplicative bias in the structure function amplitude. The actual 
value of the scale factor, which depends sensitively on the noise level, is 
difficult  to estimate from the observations. In presence of this scaling 
bias, we can estimate the power law index of the structure function. However, 
as the scale factor remains unknown, the amplitude of the measured structure 
function is only an upper limit to the true structure function.

\bsp
\label{lastpage}
\end{document}